\documentclass[number,dvips]{arxstspdf}
\usepackage{shere,flushend}
\usepackage{graphicx}

\volume{25}
\issue{2}
\pubyear{2010}
\firstpage{217}
\lastpage{226}
\doi{10.1214/10-STS333}

\makeatletter

\newtheorem{theorem}{Theorem}[section]
\newproclaim{defn}{Definition}[section]

\makeatother

\begin{document}
\begin{frontmatter}

\title{Dose Finding with Escalation with Overdose Control (EWOC) in
Cancer Clinical Trials}
\runtitle{Dose finding with Escalation with Overdose Control}

\begin{aug}
\author[a]{\fnms{Mourad} \snm{Tighiouart}\ead[label=e1]{mourad.tighiouart@emory.edu}}
\and
\author[b]{\fnms{Andr\'{e}} \snm{Rogatko}\ead[label=e2]{andre.rogatko@cshs.org}\corref{}}
\runauthor{M. Tighiouart and A. Rogatko}

\address[a]{Mourad Tighiouart is Associate Professor, Department of Biostatistics
and Bioinformatics and Winship Cancer Institute, Emory University,
Atlanta, GA 30322, USA \printead{e1}.}
\address[b]{Andr\'{e} Rogatko is Professor, Samuel Oschin Comprehensive Cancer
Institute,
Cedars-Sinai Medical Center, Los Angeles, CA 90048, USA \printead{e2}.}

\end{aug}

%
\begin{abstract}
Traditionally, the major objective in phase I trials is to identify a
working-dose
for subsequent studies, whereas the major endpoint in phase II and III
trials is
treatment efficacy. The dose sought is typically referred to as the
maximum tolerated dose
(MTD). Several statistical methodologies have been proposed to select
the MTD in cancer
phase I trials. In this manuscript, we focus on a Bayesian adaptive
design, known as
escalation with overdose control (EWOC). Several aspects of this design
are discussed,
including large sample properties of the sequence of doses selected in
the trial,
choice of prior distributions, and use of covariates. The methodology
is exemplified
with real-life examples of cancer phase I trials. In particular, we
show in the recently
completed ABR-217620 (naptumomab estafenatox) trial that omitting an
important predictor
of toxicity when dose assignments to cancer patients are determined
results in a
high percent of patients experiencing severe side effects and a
significant proportion
treated at sub-optimal doses.
\end{abstract}

%

\end{frontmatter}

\section{Introduction}

The main objective in cancer phase I clinical trials is to identify a
tolerable dose of a cytotoxic or therapeutic agent for subsequent
studies. Phase I trials represent the first testing of an
investigational agent or combination of agents whose safety profile has
been established individually. These trials typically enroll patients
with advanced cancer stages and who have exhausted available standard
treatment options \cite{Roberts2004}.

Cancer phase I trials are carried out sequentially, assigning dose
levels to subjects based on the observed side effects of the previously
treated patients. From a safety and therapeutic perspective, these
trials should be designed to minimize the number of unacceptable toxic
events and maximize the number of patients treated at an optimal dose.
Ideally, the design should control the probability of overdosing
patients at each stage of the trial, produce a sequence of doses that
converge to the MTD, and should take into account the heterogeneous
nature of cancer phase I trial patients \cite{Tighiouart2006a}.

Decisions to escalate or de-escalate dose levels in cancer phase I
trials are made after one cycle of therapy to patients. The length of a
cycle is usually between $3$ and $6$ weeks. Therefore, the target phase
I dose is typically defined in terms of treatment-related side effects,
ignoring treatment efficacy. This is due to the fact that treatment
efficacy, expressed as a reduction in tumor size or an increase in
survival, requires months (if not years) of observation \cite
{OQuigley1990,Whitehead1997}, a length of time far greater than the
length of one cycle of therapy. Thus, it can be stated that the main
objective of a cancer phase I clinical trial is to determine a safe
dose of a new drug or combination of drugs for subsequent clinical
evaluation of efficacy. This dose is known as the maximum tolerated
dose (MTD), or phase II dose. Specifically, the MTD, $\gamma$, is
defined as the dose expected to produce some degree of medically
unacceptable, dose-limiting toxicity (DLT) in a prespecified proportion
$\theta$ of patients \cite{Gatsonis1992},
%
\begin{eqnarray}
\label{defmtd}
P(\mathrm{DLT}  |  \mathrm{Dose} = \gamma) = \theta.
\end{eqnarray}

The target probability of DLT $\theta$ depends on the severity of the
treatment-attributable toxicity. It is set relatively high when the DLT
is reversible or nonfatal condition, and low if it is life-threatening
\cite{Babb1998}. Ting \cite{Ting2006} and Rosenberger and Haines
\cite{Rosenberger2002} gave good reviews of statistical methods for
dose finding in cancer phase I trials. In particular, the widely used
continual reassessment method (CRM) proposed by O'Quigley, Pepe and Fisher  \cite
{OQuigley1990} and its extensions by Faries \cite{Faries1994},
Goodman, Zahurak and Piantadosi \cite{Goodman1995}, M\"{o}ller \cite{Moller1995},
Piantadosi, Fisher and Grossman \cite{Piantadosi1998}, and Storer \cite
{Storer2001}, and the escalation with overdose control (EWOC) method
proposed by Babb, Rogatko and Zacks \cite{Babb1998}, Zacks, Rogatko and  Babb \cite
{Zacks1998}, Babb and Rogatko \cite{Babb2001}, Tighiouart, Rogatko and  Babb
\cite{Tighiouart2005} and Rogatko et al. \cite{Rogatko2008} are
Bayesian adaptive and produce consistent sequences of doses under some
model assumptions and regularity conditions. These designs can be
easily implemented in practice using published tutorials and free
interactive software; see, for example, the works of Garrett \cite
{Garrett-Mayer2006}, Zohar et al. \cite{Zohar2003}, Xu, Tighiouart and Rogatko \cite
{Xu2007}, and Rogatko, Tighiouart and Xu \cite{Rogatko2005}.

In this article, we review several aspects of EWOC, including large
sample properties, choice of prior distributions, and use of
covariates. The methodology is exemplified with cancer phase I clinical
trials we designed and conducted at Fox Chase Cancer Center in
Philadelphia and the Winship Cancer Institute in Atlanta.

This article is organized as follows. In Section~\ref{sec2}, we introduce the
phase I design known as EWOC and review its large sample properties. We
illustrate its implementation using a real-life example. An extension
of this design to account for patients' specific characteristics is
described in Section~\ref{sec3} and the methodology is illustrated by a recently
completed phase I cancer trial. Section~\ref{sec4} contains some concluding
remarks and discussion.

\section{Escalation with Overdose Control}\label{sec2}

Denote by $Y$ the binary indicator of DLT for a patient given dose $x$.
Assume that there exist $x^{*}$ and $x^{**}$, $x^{*} < x^{**}$ such that
%
\begin{eqnarray}
\label{condition1}
P(Y=1 | x=x^{*}) & = & 0,
\\ \label{eq3}
P(Y=1 | x=x^{**})& = & 1-\varepsilon,
\end{eqnarray}
 where $0 < \varepsilon< 1$ is known and $\theta< 1 - \varepsilon$.

Let $F(z)$ be a strictly increasing cumulative distribution function
(c.d.f.) having probability density function $f(z)$. We consider a
dose-toxicity relationship of the form
\begin{eqnarray}
\label{modelgeneral}
&&P(Y=1 | x)\nonumber
\\[-8pt]\\[-8pt]
&&\quad=F\biggl( F^{-1}(1-\varepsilon)+\beta\log\biggl(\frac
{x-x^{*}}{x^{**}-x^{*}}\biggr)\biggr),\nonumber
\end{eqnarray}
 where $\beta$ is unknown, and $ 0 < \beta^{*} \leq\beta
\leq
\beta^{**}$ for some positive real numbers $\beta^{*}$ and $\beta
^{**}$. This model assumes that the quantiles of $F$ are linear in the
log-standardized dose $z=\log[(x-x^{*})/(x^{**}-x^{*})]$.
An example of $F$ that is commonly used in practice is the logistic
model $F(z)=e^{z}/(1+e^{z})$. It is easy to verify that model (\ref
{modelgeneral}) satisfies the constraints (\ref{condition1}) and (\ref{eq3}).
The condition $\beta> 0$ implies that the probability of DLT is an
increasing function of dose. Let $\phi= F^{-1}(1-\varepsilon
)-F^{-1}(\theta)$. Using (\ref{modelgeneral}), it can be shown that
the MTD $\gamma$ defined in (\ref{defmtd}) is
%
\begin{eqnarray}
\label{mtd}
\gamma= x^{*}+(x^{**}-x^{*})e^{-\phi/\beta}.
\end{eqnarray}
 This also shows that $\gamma\in[x^{*}, x^{**}]$. Let
$\gamma
'=\log((\gamma-x^{*})/(x^{**}-x^{*}))$ be the MTD on the
log-standardized scale. Then (\ref{mtd}) implies that $\gamma' = -\phi/\beta$.

\subsection{Dose Escalation Based on Bayesian Estimates}\label{sec2.1}

Let $G(u) = F( F^{-1}(\theta) + \phi+ u), g(u) = G'(u)$ and $z_{1} =
-\phi/ \beta^{*}$ be the level assigned to the first patient. Then,
\begin{eqnarray*}
G(\beta z_{1})&=& F\bigl(F^{-1}(\theta)+F^{-1}(1-\varepsilon)-F^{-1}(\theta
)+\beta z_{1}\bigr)
\\
&=&F\bigl(F^{-1}(1-\varepsilon)+\beta z_{1}\bigr)
\\
&=&P(Y=1|z_{1}) \leq F\bigl(F^{-1}(1-\varepsilon)+\beta^{*}
z_{1}\bigr)
\\
&=&F(F^{-1}(\theta))=\theta,
\end{eqnarray*}
 since $z_{1}<0$ and $F(z)$ is strictly increasing. This
shows that
this log-standardized dose $z_{1}$ is safe in the sense that the
probability of DLT at this level does not exceed $\theta$. Let
$D_{n}=\{( z_{i}, Y_{i}), i = 1,\ldots,n\}$ be the data after
enrolling $n$ patients to the trial where $Y_{i}$ is the observed DLT
status of the patient getting level $z_{i}$, $z_{i}\in L^{*}=[-\frac
{\phi}{\beta^{*}}, -\frac{\phi}{\beta^{**}}]$.


Let $h(\beta)$ be a prior density function for the parameter $\beta$
on $[\beta^{*} ,\beta^{ **}]$ and $\Pi_{n}(\beta)=\Pi(\beta
|D_{n})$ the posterior c.d.f. given the data $D_{n}$. Let $0 < \alpha
<1$. A sequence of dose levels $z_{n}$ such that
%
\begin{eqnarray}
\label{feasible}
P(z_{n} \leq-\phi/ \beta|D_{n-1}) \geq1-\alpha
\end{eqnarray}
 for all $n \geq2$ is called Bayesian-feasible at level $(1-
\alpha
)$; see the article by Zacks, Rogatko and Babb \cite{Zacks1998}. Let
%
\begin{eqnarray}
\label{xialpha}
z_{n}^{(\alpha)}=-\frac{\phi}{\Pi_{n-1}^{-1}(\alpha)}, n \geq2.
\end{eqnarray}
 Then, it is easy to verify that for all $n \geq2$,
$z_{n}^{(\alpha
)}$ is Bayesian-feasible at level $(1- \alpha)$. The choice of
$z_{n}^{(\alpha)}$ as the log-standardized dose levels in the trial
implies that the posterior probability of exceeding the MTD is equal to
the feasibility bound $\alpha$. Let $\mathcal{F}_{n} = \sigma
(D_{n})$ be the sigma-field generated by $D_{n}$ and $\psi^{(\alpha
)}$ be the class of all Bayesian-feasible sequences $z_{n} \in\mathcal
{F}_{n}$ of level $(1- \alpha)$.

\begin{defn}
A sequence of levels $\{z_{n}^{*}, n \geq1 \} \in\psi^{(\alpha)}$
is called optimal Bayesian-feasible at level $(1- \alpha)$, if for all
$N \geq1$,
\[
\sum_{n=1}^{N} E_{h}\{(\gamma'-z_{n}^{*})^{+}\}=\inf_{\{z_{n}\}\in
\psi^{(\alpha)}} \sum_{n=1}^{N} E_{h}\{(\gamma'-z_{n})^{+}\},
\]
 where $z^{+} = z I(z > 0)$ denotes the positive part of a
random variable.
\end{defn}

This means that $z_{n}^{*}$ minimizes the average amount by which
patients are underdosed. Using the law of total expectation, Zacks, Rogatko and Babb \cite{Zacks1998}
showed that $z_{n}^{(\alpha)}$ is optimal
Bayesian-feasible. Conditions under which this sequence converges to
the true MTD in probability are stated in the next theorem.

\begin{theorem}
\label{theorem1}
Suppose that for $\beta_{0} \in[\beta^{*},\beta^{**}]$:
\begin{enumerate}[1.]

\item$0<\varepsilon_{1}<G(-\beta_{0} \phi/\beta^{*})\leq G(-\beta
_{0} \phi/\beta^{**})\leq1-\varepsilon$.

\item$0<\varepsilon_{2}<\inf\{g(\beta_{0}x)\dvtx x\in L^{*}\}\le\sup\{
g(\beta_{0}x)\dvtx x\in L^{*}\} \leq g^{*}$.

\item$g(x)$ is continuously differentiable.

\item$-\infty< \inf\{g'(\beta_{0}x)\dvtx x\in L^{*}\}\le\sup\{g'(\beta
_{0}x)\dvtx x\in\break L^{*}\}<\infty$.
\item$h(\beta)$ is uniform on $[\beta^{*} , \beta^{**}]$. Then,
$z_{n}^{(\alpha)} \stackrel{p}{\longrightarrow} -\phi/\break \beta_{0}$
as $n \rightarrow\infty.$
\end{enumerate}
\end{theorem}

\begin{pf}
See the article by Zacks, Rogatko and Babb \cite{Zacks1998}.
\end{pf}

\subsection{Coherence of EWOC}\label{sec2.2}

Coherence of adaptive designs was introduced by Cheung \cite
{Cheung2005} in the context of cancer phase I clinical trials. Due to
ethical concerns, the dose of a cytotoxic agent for the next patient in
a trial should not be higher than the current allocated dose if the
current patient exhibits DLT. Likewise, the dose for the next patient
should not be lower than the current one if the current patient does
not exhibit DLT. This desirable property is known as coherence and
Cheung \cite{Cheung2005} showed that CRM is coherent. The author also
showed how the coherence property can be lost when ad hoc modifications
are introduced to CRM. In this section, we show that EWOC as described
in Section~\ref{sec2.2} is coherent.

Let $F(x,\gamma) = P(Y=1|x)$ be the model given in (\ref
{modelgeneral}) reparameterized in terms of the MTD $\gamma$. Let
$D_{n}=\{(x_{1},Y_{1}),\ldots,(x_{n},Y_{n})\}$ be the data generated
using the EWOC scheme described in Section~\ref{sec2.2}. This design is said to
be coherent in escalation if for all $n \geq2, x_{n} \geq x_{n-1}$
whenever $Y_{n-1} = 0$. The design is said to be coherent in
de-escalation if for all $n \geq2, x_{n} \leq x_{n-1}$ whenever
$Y_{n-1} = 1$. The design is said to be coherent if it is coherent in
both escalation and de-escalation.

\begin{theorem}
\label{coherthm}
Suppose that $F(x,\gamma)$ is nonincreasing in $gamma$ for fixed dose
$x$. Then the EWOC scheme described in Section~\ref{sec2.2} is coherent.
\end{theorem}

The proof of Theorem \ref{coherthm} is given in the \hyperref[app]{Appendix}. It is
easy to verify that the monotonicity condition on $F(x,\gamma)$ is
satisfied by model (\ref{modelgeneral}), and in particular, the
logistic function.

\subsection{Two-Parameter Logistic Model}\label{sec2.3}

Denote by $X_{\min}$ and $X_{\max}$ the minimum and maximum dose
levels available for use in the trial. One chooses these levels in the
belief that $X_{\min}$ is safe when administered to humans.
Babb, Rogatko and Zacks \cite{Babb1998} considered a two-parameter logistic model for the
dose-toxicity relationship:
%
\begin{eqnarray}
\label{twoparam}
P(Y=1|\mathit{Dose} = x) = \frac{\exp(\beta_{0} + \beta_{1}x)}{1+ \exp
(\beta_{0} + \beta_{1}x)},
\end{eqnarray}
 where we assume that $\beta_{1}>0$ so that the probability
of DLT
is a monotonic increasing function of dose. Model (\ref{twoparam}) is
reparameterized in terms of the MTD $\gamma$ and the probability of
DLT at the starting dose $\rho_{0}$, parameters clinicians can easily
interpret. This might be advantageous since $\gamma$ is the parameter
of interest and one often conducts preliminary studies at or near the
starting dose so that one can select a meaningful informative prior for
$\rho_{0}$. Using the definition of the MTD in (\ref{defmtd}) and
(\ref{twoparam}), it can be shown that

\begin{eqnarray}
\label{reparam}
\beta_{0} & = & \frac{X_{\min} \rm \operatorname{logit}(\theta) - \gamma
\operatorname{logit}(\rho_{0})}{x_{\min}-\gamma},\nonumber
\\[-8pt]\\[-8pt]
\beta_{1} & = & \frac{\operatorname{logit}(\rho_{0})-  \operatorname{logit}(\theta
)}{x_{\min}-\gamma}. \nonumber
\end{eqnarray}

The second equation in (\ref{reparam}) shows that the assumption that
$\beta_{1} > 0$ implies $0 < \rho_{0} < \theta$.

\subsubsection{Trial design}

After specifying a prior distribution $h(\rho_{0},\gamma)$ for $(\rho
_{0},\gamma)$, denote by $\Pi_{n}(\gamma)$ the\break marginal posterior
c.d.f. of $\gamma$ given $D_{n}$. EWOC can be described as follows.
The first patient receives the dose $x_{1} = X_{\min}$ and conditional
on the event $\{y_{1} = 0\}$, the $(n+1)$st patient receives the dose
$x_{n+1}= \Pi_{n}^{-1}(\alpha)$ so that the posterior probability of
exceeding the MTD is equal to the feasibility bound $\alpha$. If
$y_{1} = 1$, we recommend that the clinician stops the trial.
Calculation of the marginal posterior distribution of $\gamma$ is
performed using numerical integration; see \cite{Babb1998}. Often in
practice, phase I clinical trials are typically based on a small number
of prespecified dose levels $d_{1}, \ldots,d_{r}$. In this case, the
$(n+1)$st patient receives the dose
\begin{eqnarray*}
\hat{d}_{n+1}&=&\max_{1\leq i \leq r}\{d_{i} \dvtx d_{i}-x_{n+1} \leq T_{1}
  \\
  &&\qquad\hspace*{4pt} \mbox{and }    \Pi_{n}(x_{n+1})-\alpha\leq T_{2}\},
\end{eqnarray*}
where $T_{1}, T_{2}$ are nonnegative numbers we refer to as
tolerances. We note that this design scheme does not require that we
know all patient responses before we can treat a newly accrued patient.
Instead, we can select the dose for the new patient on the basis of the
data currently available.
At the conclusion of the trial, the MTD is estimated by minimizing the
posterior expected loss with respect to some suitable loss function~$l$.
One should consider asymmetric loss functions since
underestimation and overestimation have very different consequences.
Indeed, the dose $x_{n}$ selected by EWOC for the $n$th patient
corresponds to the estimate of $\gamma$ having minimal risk with
respect to the asymmetric loss function
\[
l_{\alpha}(x,y)= \cases{
\alpha(\gamma- x),
\cr\quad\mbox{if $x \leq\gamma$, that is, if $x$ is an
underdose},
\cr
(1-\alpha)(x - \gamma),
\cr\quad\mbox{if $ x> \gamma$, that is, if $x$ is
an overdose}.
}
\]

Note that the loss function $l_{\alpha}$ implies that for any $\delta
> 0$, the loss incurred by treating a patient at $\delta$ units above
the MTD is $(1 - \alpha) / \alpha$ times greater than the loss
associated with treating the patient at $\delta$ units below the MTD.
This interpretation might provide a meaningful basis for the selection
of the feasibility bound. The above methodology can be implemented
using the user-friendly software of Rogatko, Tighiouart and Xu \cite{Rogatko2005}.

\subsubsection{Correlated priors on $\rho_{0}$ and $\gamma$}

In models (\ref{modelgeneral}) and (\ref{twoparam}), we assumed that
the support of the MTD was strictly contained in $[x^{*}, x^{**}]$ and
$[X_{\min}, X_{\max}]$, respectively. The assumption that $\gamma$
is bounded from above may be too restrictive. In the absence of
toxicity, this assumption causes the dose escalation rate to slow down
and in general, the target MTD will never be achieved if it lies
outside the support of $\gamma$. Furthermore, since the support of the
probability of DLT at the initial dose $\rho_{0}$ is $[0, \theta]$
and $\gamma$ is a function of $\theta$, the assumption of prior
independence between $\rho_{0}$ and $\gamma$ may not be realistic.
Intuitively, the closer $\rho_{0}$ is to $\theta$, the closer the MTD
is to $X_{\min}$. Tighiouart, Rogatko and Babb \cite{Tighiouart2005} introduced a
class of correlated priors for $h(\rho_{0}, \gamma)$ on $[0, \theta
]\times[X_{\min} , \infty)$ using truncated normal distributions for
the parameter $\gamma$. They showed that a candidate joint prior for
$(\rho_{0}, \gamma)$ with negative a priori correlation structure
results in a safer trial than the one that assumes independent priors
for these two parameters while keeping the efficiency of the estimate
of the MTD essentially unchanged.

\subsection{EWOC with Varying Feasibility Bound}

Many of the phase I cancer trials the authors designed at Fox Chase
Cancer Center and Winship Cancer Institute used a variable feasibility
bound $\alpha$; see the work of Babb and
Rogatko \cite{Babb2004,Babb2001}, Cheng et al. \cite{Cheng2004}, Tighiouart and
Rogatko \cite{Tighiouart2006a,Tighiouart2006}, and Xu, Tighiouart and Rogatko \cite
{Xu2007}. The rationale behind this approach is that uncertainty about
the MTD is high at the onset of the trial and a small value of $\alpha
$ offers protection against the possibility of administering dose
levels much greater than the MTD. As the trial progresses, uncertainty
about the MTD declines and the likelihood of selecting a dose level
significantly above the MTD becomes significantly smaller. However,
design operating characteristics were not studied. Chu, Lin and Shih \cite
{Chu2009} compared the performance of different versions of CRM with
EWOC with both constant and varying $\alpha$. The design of EWOC with
varying $\alpha$ was termed ``hybrid design.'' The authors conducted
extensive simulations to compare these designs in terms of (1) the
proportion of patients given doses above the ``true'' MTD and (2) the
proportion of times the recommended dose is the ``true'' MTD after each
patient is enrolled in the trial and his or her DLT status is resolved.
It was found in general that both the hybrid and CRM designs had better
convergence rate than EWOC with fixed $\alpha$ and that EWOC with
fixed and varying feasibility bound $\alpha$ provide a better overdose
protection than the CRM designs in the sense that fewer patients are
given doses above the ``true'' MTD.

\subsection{Example}

EWOC was used to design a phase I clinical trial that involved the
R115777 drug at Fox Chase Cancer Center in Philadelphia, USA in 1999.
R115777 is a selective nonpeptidomimetic inhibitor of
farnesyltransferase (FTase), one of several enzymes responsible for
posttranslational modification that is required for the function of
p21(ras) and other proteins. This was a repeated dose, single center
trial designed to determine the MTD of R115777 in patients with
advanced incurable cancer. The target probability of DLT was set to
$\theta= 1/3$.
The dose-escalation scheme was designed to determine the MTD of R115777
when drug is administered orally for 12 hours during 21 days followed
by a 7-day rest. This constitutes one cycle of therapy. Toxicity was
assessed by the National Cancer Institute (NCI) Common Toxicity
Criteria \cite{NCI2003}. DLT was determined by week 3 of cycle 1, as
defined by Grade III nonhematological toxicity (with the exception of
alopecia or nausea/vomiting) or hematological Grade IV toxicity with a
possible, probable or likely causal relationship to administration of
R115777. Dosing continued until there was evidence of tumor progression
or DLT leading to permanent discontinuation. The initial dose judged to
be safe by the clinician for this study was $X_{\min} = 60$~$\rm
mg/\rm m^{2}$ and the maximum allowable dose was $X_{\max} = 600$~$\rm
mg/\rm m^{2}$. More details about the dosing regimen for this trial can
be found in the work of Tighiouart and Rogatko \cite
{Tighiouart2006a}. Assuming vague priors for $\rho_{0}$ on $[0, \theta
]$ and $\gamma$ on $[60, 600]$, the prior probability density of
$(\rho_{0},\gamma)$ is
\begin{eqnarray*}
&&h(\rho_{0},\gamma)
\\
&&\quad= \cases{
1/180, &\mbox{if $(\rho_{0},\gamma) \in[0,1/3] \times[60, 600]$},
\cr
0, &\mbox{otherwise}.
}
\end{eqnarray*}

\begin{figure}

\includegraphics{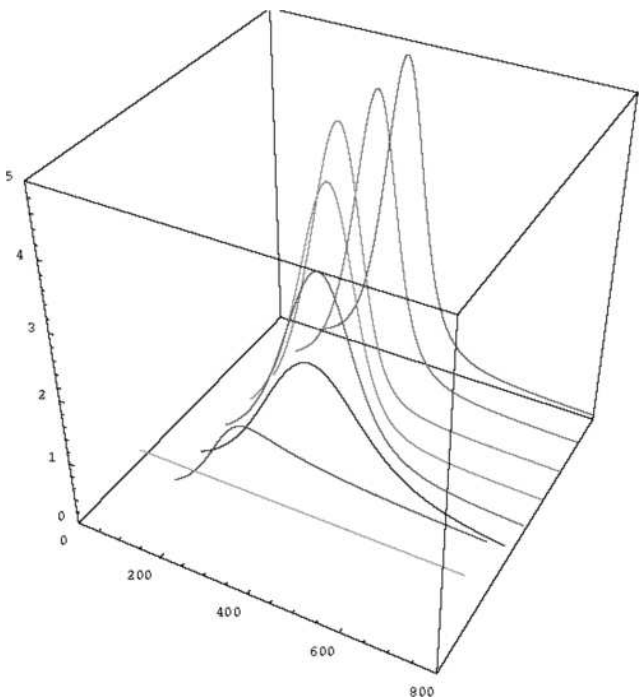}

\caption{Posterior density of the MTD when the number of treated patients
(from bottom to top) is $1, 5, 10, 15, 20, 25, 30, 33$.}
\label{fig1}
\end{figure}

\begin{figure}

\includegraphics{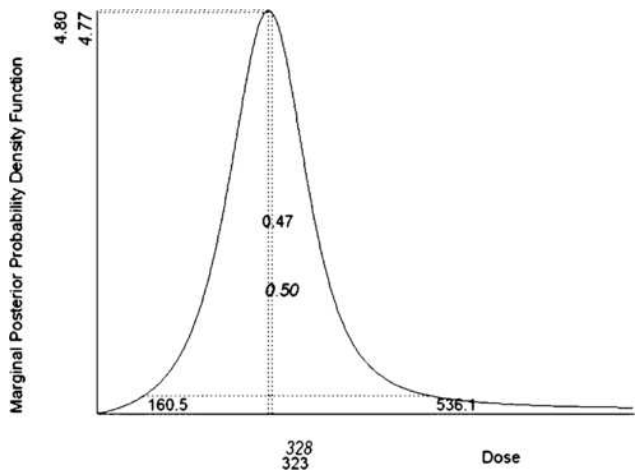}

\caption{Posterior density of the MTD after $33$ patients have
been treated. The posterior mode is $323$ ($47$th percentile)
and the median and dose to be given to the $34$th patient is $328$.
The $95\%$ highest posterior density interval is $[160.5, 536.1]$.}
\label{fig2}
\end{figure}

Thus, $\rho_{0}$ and $\gamma$ are independent a priori, uniformly
distributed over their corresponding interval. Figure \ref{fig1} shows
the posterior distributions of the MTD as the trial progressed and
Figure \ref{fig2} shows the posterior density of the MTD after $33$
patients have been treated. The posterior mode is $323$ which
corresponds to the $47$th percentile of the distribution. In this
trial, we used a variable feasibility bound $\alpha$, starting with
$\alpha= 0.3$, this value being a compromise between the therapeutic
aspect of the agent and its toxic side effects. As the trial
progressed, $\alpha$ increased in small increments until $\alpha=
0.5$ so that, by the end of the trial, the given dose corresponds to
the $50$th percentile, that is, the median of the marginal posterior
probability density function. Thus, the dose to be given to the $34$th
patient is $328$. The $95$\% highest posterior density interval is
$[160.5, 536.1]$.

\section{Adjusting For Patients' Baseline Covariates}\label{sec3}

A key assumption implied by the definition of the phase I target dose
(MTD) is that every subgroup of the patient population has the same
MTD. That is, it is assumed that the patient population is homogeneous
in terms of treatment tolerance and every patient should be treated at
the same dose. As a result, no allowance is made for individual patient
differences in susceptibility to treatment \cite{Dillman1992}.

Babb and Rogatko \cite{Babb2001} extended EWOC to allow the
incorporation of information concerning individual patient differences
in susceptibility to treatment. The method adjusts doses according to
patient-specific characteristics while safeguarding against overdosing.

\subsection{Model}

Let $W$ be a $p$-dimensional baseline covariate vector. We consider the
dose-toxicity model
\begin{eqnarray}
\label{covmodel}
&& P(Y=1|\mathit{Dose} = x, W= w)\nonumber
\\[-8pt]\\[-8pt]
&&\quad = \frac{\exp(\beta_{0} + \beta_{1}x + \eta'
w)}{1+ \exp(\beta_{0} + \beta_{1}x +\eta' w)},\nonumber
\end{eqnarray}
 where $\eta\in\mathbb{R}^{p}$ is the effect of the baseline
covariate vector on DLT. Let $p_{x}(w)=P(Y=1|\mathit{Dose} = x, W= w)$. We
assume that $\beta_{1}>0$ so that $p_{x}(w)$ is an increasing function
of dose $x$ for fixed $w$. The MTD for a patient with baseline
covariate value $w$ is defined as the dose $\gamma(w)$ that results in
a probability equal to $\theta$ that a DLT will manifest. It follows
from model (\ref{covmodel}) that
%
\begin{eqnarray}
\label{condmtd}
\gamma(w)=\beta_{1}^{-1}\biggl[\log\biggl(\frac{\theta}{1-\theta
}\biggr)-\beta_{0}-\eta'w\biggr].
\end{eqnarray}

As in Section~\ref{sec2.3}, we reparameterize this model in terms of $(\gamma
(w^{*}),\rho)$ for a selected value of the baseline covariate vector
$w=w^{*}$ and $\rho$ is a ($p+1$)-dimensional nuisance parameter.

\subsubsection{Trial design}

Let $h(\rho, \gamma(w^{*}))$ be a prior distribution for $(\rho
,\gamma(w^{*}))$ and denote by $\Pi_{n,w^{*}}(\gamma(w^{*}))$ the
marginal posterior c.d.f. of $\gamma(w^{*})$ given the data $D_{n}=\{
(x_{1},Y_{1},w_{1}),\dots, (x_{n},Y_{n},w_{n})\}$. The first patient
receives the dose $x_{1} = X_{\min}$ and conditional on the event $\{
y_{1} = 0\}$, the $(n+1)$st patient with covariate vector value
$w_{n+1}$ receives the dose $x_{n+1}=\break \Pi_{n,w_{n+1}}^{-1}(\alpha)$
so that the posterior probability of exceeding the MTD is equal to the
feasibility bound $\alpha$. Note that here, $\Pi_{n,w_{n+1}}^{-1}(\cdot)$
is the inverse c.d.f. of $\Pi_{n,w_{n+1}}(\gamma(w_{n+1}))$.

For a binary covariate $W$, Tighiouart, Rogatko and Xu \cite{Tighiouart2007}
studied operating characteristics of this model with extensive
simulations under different scenarios for the underlying true MTDs.
They found that if the two MTDs are different and the design does not
adjust for this heterogeneity, then the trial will result in more
patients being overdosed. If the two MTDs are different and parallel
trials are used, then the estimates of the MTDs are less efficient.
Finally, if the two MTDs are the same and the design adjusts for
patients' heterogeneity, then few more patients can be overdosed if the
true MTD is low relative to a design with no covariate but the
difference is not practically important. Thus, we stand to lose little
if we do include a statistically nonsignificant covariate in the model.
This conclusion is in agreement with the findings of O'Quigley, Shen and Gamst
\cite{OQuigley1999}.

\subsection{Example}\label{sec3.2}

ABR-217620 (naptumomab estafenatox) is a recombinant fusion protein
that consists of the 5T4Fab moiety genetically fused to the engineered
superantigen variant SEA/E-120. This fusion protein is a new generation
tumor-targeted superantigen based on the previously described
ABR-214936\break (anatumomab mafenatox). ABR-217620 was\break designed to reduce
antigenicity and toxicity. We use model (\ref{covmodel}) with $W=C$
representing the Anti\break SEA/E120 covariate to design a phase I study for
nonsmall cell lung cancer (NSCLC) patients. The goal is to determine
the MTD of ABR-217620 as a function of patients' baseline Anti SEA/E120
and test whether the neutralizing effect of Anti SEA/\break E120 on the
cytotoxic agent which was observed by Babb and Rogatko \cite
{Babb2001} has been reduced or eliminated with this new agent. The
modeling approach is similar to the PNU trial described in \cite
{Babb2001}, the target probability of DLT $\theta$ was set to $0.2$.
The feasibility bound $\alpha$ was set at 0.25 for the first nine
patients, then was increased to a maximum value of $0.5$ by increments
of $0.05$ every time a new patient was enrolled in the trial and a DLT
assessment was resolved. Based on preliminary clinical data, the
minimum and maximum allowable doses for ABR-217620 set by the
clinicians are $x_{\min} = 1$ ug/kg and $x_{\max} = 100$ ug/kgl. The
minimum and maximum values of Anti SEA/E120 anticipated in the trial
are $c_{1} = 0$ pmol/ml and $c_{2} = 200$ pmol/ml, respectively. As in
the PNU trial \cite{Babb2001}, we reparameterize model (\ref
{covmodel}) in terms of $\gamma_{\max}=\gamma(c_{2}), \rho
_{1}=p_{x_{\min}}(c_{1}), \rho_{2}=p_{x_{\min}}(c_{2})$ with
$(\gamma_{\max}, \rho_{1}, \rho_{2})$ uniformly distributed on $\{
(x, y, z)\dvtx y \in(0,\theta], z \in(0, y), x \in[1,100]\}$ a priori.

\begin{figure}

\includegraphics{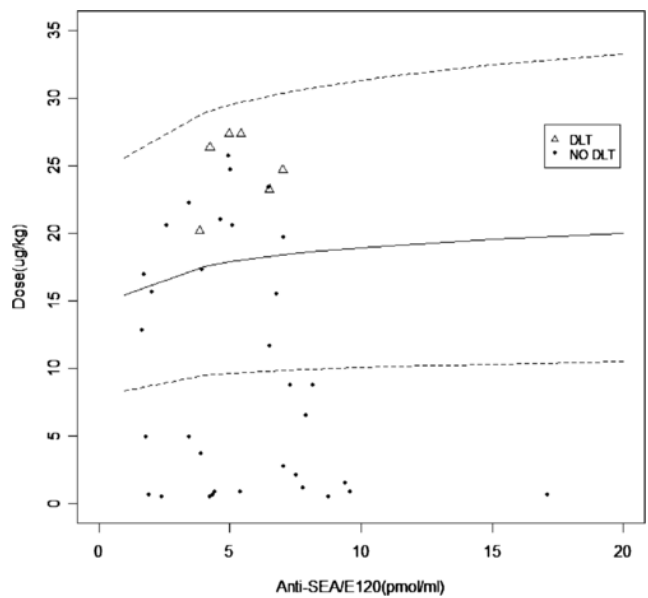}

\caption{Dose allocation as a function of baseline Anti SEA/E120 during
the trial for all $39$ patients. The solid line is the MTD conditional on
the covariate Anti SEA/E120 which corresponds to the posterior median of
the conditional posterior distribution of the MTD and the dashed lines
delimit the $95\%$ Bayesian credible region.}\vspace*{-8pt}
\label{fig3}
\end{figure}
\begin{figure} 

\includegraphics{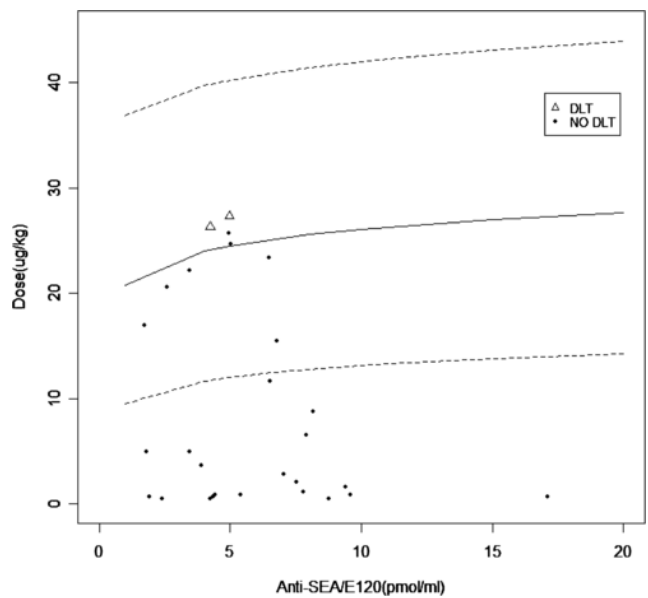}

\caption{Dose allocation as a function of baseline Anti SEA/E120 during the
trial for all $28$ NSCLC \& PC patients. The solid line is the MTD conditional
on the covariate Anti SEA/E120 which corresponds to the posterior median of
the conditional posterior distribution of the MTD and the dashed lines delimit
the $95\%$ Bayesian credible region.}
\label{fig4}
\end{figure}

\begin{figure}[t]

\includegraphics{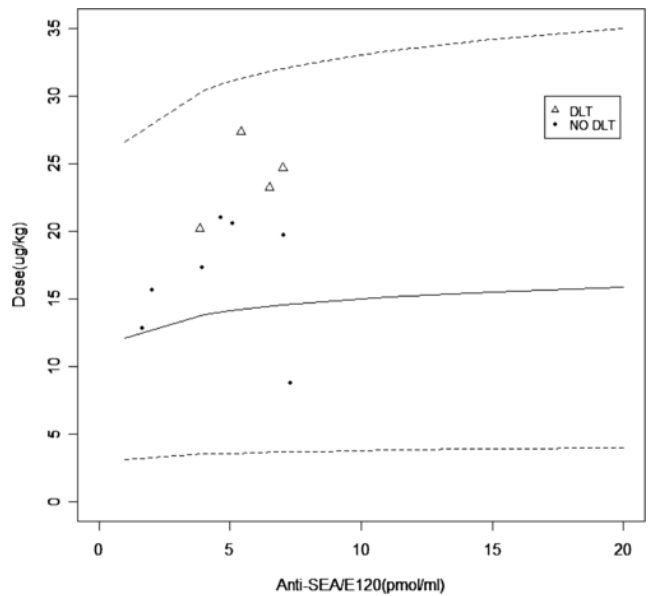}

\caption{Dose allocation as a function of baseline Anti SEA/E120 during the
trial for all $11$ RCC patients. The solid line is the MTD conditional on the
covariate Anti SEA/E120 which corresponds to the posterior median of the
conditional posterior distribution of the MTD and the dashed lines delimit the
$95\%$ Bayesian credible region.}
\label{fig5}
\end{figure}

Figure \ref{fig3} shows the doses allocated to all 39 patients as a
function of their pretreatment Anti\break SEA/E120. The solid line is the
estimated conditional MTD, obtained by taking the posterior median of
the marginal distribution of the MTD conditional on the covariate Anti
SEA/E120. The dashed lines delimit the $95\%$ Bayesian credible region.
Six patients experience DLT ($15.4\%$) and the MTD seems to indicate
that the neutralizing effect of Anti\break SEA/E120 has been reduced
considerably. The protocol was amended to include patients with renal
cell (RCC) and pancreatic cancer (PC). Figures \ref{fig4} and \ref
{fig5} show the doses allocated to NSCLC $\&$ PC patients and RCC
patients, respectively. The solid line represents the conditional MTD
obtained after fitting the data in each group to model (\ref
{covmodel}), reparameterized in terms of $(\gamma_{\max}, \rho_{1},
\rho_{2})$. This shows that NSCLC $\&$ PC patients were treated at
sub-optimal doses and RCC patients were overdosed, with $36.4\%$
experiencing DLT, way above the target probability of DLT $\theta=0.2$.

The effects of Anti SEA/E120 and type of cancer were tested by fitting
model (\ref{covmodel}) with $W=(C,Z)$, where $C$ is the baseline Anti
SEA/E120 and $Z$ is a binary covariate representing the cancer type, $Z
= z_{1} = 1$ for NSCLC and PC patients and $Z = z_{2} = 0$ for RCC
patients. To be consistent with the priors used to design the trial, we
reparameterized the model in terms of $\gamma_{\max}=\gamma
(c_{2},z_{1}), \rho_{1}=p_{x_{\min}}(c_{1},z_{1}),\break \rho
_{2}=p_{x_{\min}}(c_{2},z_{1})$ and $\rho_{3}=p_{x_{\min
}}(c_{1},z_{2})$. Independent uniform priors are placed on these
parameters. It can be shown that this induces priors centered at $0$
for the Anti SEA/E120 and cancer type effect parameters $\eta_{1}$ and
$\eta_{2}$. We used WinBUGS \cite{Lunn2000} to fit this model and
the $95\%$ HPD intervals for the parameters $\eta_{1}$ and $\eta_{2}$
were $(-0.14, 0.24)$ and $(-4.6,0.6)$, respectively. We conclude that
the agent ABR-217620 was successful in reducing the neutralizing
capacity of Anti SEA/E120 and that the phase II dose should be
carefully tailored to account for patients' cancer type and hence avoid
excessive overdosing and underdosing patients.

\section{Discussion}\label{sec4}

In this article, we described EWOC, a Bayesian dose finding design for
cancer phase I clinical trials. The method is flexible enough to allow
prior information about the drug from laboratory or animal studies to
be incorporated in the model, is coherent, makes use of all the
information available at the time of each dose assignment and controls
the probability of overdosing patients at each stage. EWOC can be
implemented with the user-friendly software EWOC 2.1 \cite
{Rogatko2005} or WinBUGS \cite{Lunn2000} for general class of prior
distributions \cite{Tighiouart2005}. The two-parameter model
described in Section~\ref{sec2.3} accounts for the uncertainty regarding the
probability of DLT at the initial dose by placing a vague prior
distribution on $\rho_{0}$. If expert opinion about this parameter is
available, then it should be incorporated in the prior for $\rho_{0}$.
In particular, if the clinician strongly believes that this prior can
be approximated by a point mass distribution, then the one-parameter
model described in Section~\ref{sec2.1} may be used. In any case, design
operating characteristics should be performed with a sensitivity
analysis about the parameter $\rho_{0}$ when designing the trial. Our
own experience in designing dose-finding studies in cancer is that the
uncertainty of the clinicians regarding the probability of DLT at the
initial dose is large. Thus, in more than ten years of designing trials
with EWOC, the use of a one-parameter model was never chosen by the
clinical researchers we worked with.

It is worth highlighting that the values of $\alpha$ and $\theta$ are
chosen independently when the trial is designed. They have distinct
meanings and functions. For example, taking a value of $\alpha$
greater than $\theta$ only affects the loss function used to estimate
the next dose and the MTD at the conclusion of the trial. It does not
mean that patients are given doses at a rate above the target
probability of DLT $\theta$. When $\alpha= 0.5$, the method differs
from CRM in the sense that the loss functions are different. The loss
function for EWOC is taken with respect to the parameter $\gamma$, the
MTD. The overprotection property of EWOC is with respect to the
posterior distribution of the MTD, given the data. The overprotection
property states that the posterior probability of exceeding the MTD
given the current data is bounded by $\alpha$. This overprotection is
as good as the posterior distribution of the MTD at each stage of the
trial. For instance, if we used a flat prior on the MTD and the true
MTD turns out to be very close to the initial dose, then it would take
many patients for the median of the posterior distribution to cluster
around the true MTD.

Another aspect of cancer phase I clinical trials not discussed here is
the choice of the number of patients to enroll. Most sample size
recommendations in the literature are based on prespecified stopping
rules; see, for example, the work of Zohar and Chevret \cite
{Zohar2001} on selecting the number of patients by considering
different stopping rules using the CRM. Lin and Shih \cite{Lin2001}
and Ivanova \cite{Ivanova2006} described sample size recommendations
based on the expected number of patients allocated to each dose
selected from a set of prespecified dose levels. However, these methods
apply to a prespecified set of discrete doses and it is not clear how
they can be applied to continuous doses. Unlike the frequentist
approach, there is no consensus on a specific Bayesian method for the
sample size determination problem; see the article by Adcock \cite
{Adcock1997} for a review of Bayesian approaches. We conducted
extensive simulation studies in order to estimate the sample size based
on a desired accuracy of the Bayes estimate on the average.
Specifically, we determined the minimum number of patients so that the
posterior variance of the MTD on the average over all possible trials
is no more than a specified margin. Tabulated values of the average
mean posterior standard deviation, length of $90\%$ and $95\%$ HPD
intervals for different values of the target probability of DLT $\theta
$ are available from the authors upon request.

The methodology described in this article assumes that DLT status is
binary and does not account for patients' time to toxicity. Information
on time to DLT is crucial to clinicians in that it permits a dynamic
updating of the posterior distribution of the MTD based on the number
of patients who experienced DLT and the ones who are still at risk. If
new patients are eligible to enter the clinical trial while the DLT
status of currently enrolled patients is still being resolved, then the
new patients are allocated to the current established dose because it
is not ethical to resolve DLT status at the expense of treatment delay.
In this case, there is no adaptation to the most current information
and the design will not be efficient. Time to DLT was first
investigated by Cheung and Chapell \cite{Cheung2000} and later
adapted to estimation of a maximum cumulative dose by Braun et al.
\cite{Braun2003}. These methods are extensions of the CRM and
incorporate information on partially observed patients using weighted
binomial likelihoods. EWOC can be adapted to this framework by modeling
time to DLT as a Cox \cite{Cox1972} type model with a parametric or
nonparametric baseline risk of toxicity $h_{0}(t)$. We are currently
investigating the performance of a large class of models within this
framework via extensive simulations.

\appendix
\section*{Appendix}\label{app}

\begin{pf*}{Proof of Theorem \ref{coherthm}}
Let $\Pi_{n}(t)$ be the posterior c.d.f. of $\gamma$ given $D_{n}, n
\geq2$. Then, it suffices to show that:
\begin{enumerate}[1.]
\item$\Pi_{n}(t) \leq\Pi_{n-1}(t)$ for all $t$ whenever $Y_{n}=0$.

\item$\Pi_{n}(t) \geq\Pi_{n-1}(t)$ for all $t$ whenever $Y_{n}=1$.
\end{enumerate}

Let
\[
L_{n}(\gamma|D_{n})=\prod_{i=1}^{n} (F(x_{i},\gamma))^{Y_{i}} \bigl(1
-F(x_{i},\gamma)\bigr)^{1-Y_{i}}
\]
 be the likelihood function and $h(\gamma)$ be a proper prior
distribution for $\gamma$. To simplify notation, let $L_{n}(\gamma
|D_{n})=L_{n}(\gamma)$, $F_{i}(\gamma)=F(x_{i},\gamma)$ and suppose
that $x^{*}=0, x^{**}=1$. Using Bayes' rule, the posterior c.d.f. $\Pi
_{n}(t)$ given $D_{n}$ is
\[
\Pi_{n}(t)=\frac{\int_{0}^{t}L_{n}(\gamma) h(\gamma)\,d\gamma}{\int
_{0}^{1} L_{n}(\gamma) h(\gamma) \,d\gamma}.
\]

Suppose that $Y_{n}=0$. Then, $L_{n}(\gamma)=L_{n-1}(\gamma)
(1-F_{n}(\gamma))$ and
\[
\Pi_{n}(t)=\frac{\int_{0}^{t}L_{n-1}(\gamma) (1-F_{n}(\gamma))
h(\gamma)\,d\gamma}{\int_{0}^{1} L_{n-1}(\gamma)(1-F_{n}(\gamma))
h(\gamma) \,d\gamma}.
\]

It follows that
\begin{eqnarray*}
&&\Pi_{n}(t)-\Pi_{n-1}(t)
\\
&&\quad =\frac{\int_{0}^{t} L_{n-1}(\gamma)
(1-F_{n}(\gamma)) h(\gamma)\, d\gamma}{\int_{0}^{1} L_{n-1}(\gamma
)(1-F_{n}(\gamma)) h(\gamma)\, d\gamma}
\\
&&\qquad{}- \frac{\int_{0}^{t}
L_{n-1}(\gamma) h(\gamma)\, d\gamma}{\int_{0}^{1} L_{n-1}(\gamma)
h(\gamma) \,d\gamma}
\\
&&\quad{}=A^{-1} \biggl[\int_{0}^{t}\hspace*{-2pt}\int_{0}^{1} L_{n-1}(\gamma
)L_{n-1}(\gamma') h(\gamma)h(\gamma')
\\
&&{}\hspace*{57pt}\qquad\times[F_{n}(\gamma')-F_{n}(\gamma
)]\,d\gamma'\, d\gamma\biggr],
\end{eqnarray*}
 where
\begin{eqnarray*}
A &=& \int_{0}^{1}\hspace*{-3pt}\int_{0}^{1} L_{n-1}(\gamma)\bigl(1-F_{n}(\gamma
)\bigr)h(\gamma)
\\
&&L_{n-1}(\gamma')h(\gamma')\,d\gamma'\, d\gamma,
\end{eqnarray*}
\begin{eqnarray*}
&&\Pi_{n}(t)-\Pi_{n-1}(t)
\\
&&\quad=A^{-1} \int_{0}^{t}\hspace*{-3pt}\int_{0}^{t}
L_{n-1}(\gamma)L_{n-1}(\gamma') h(\gamma)h(\gamma')
\\
&&\qquad{}\hspace*{44pt}\times[F_{n}(\gamma
')-F_{n}(\gamma)]\,d\gamma'\, d\gamma
\\
&&{}\qquad + A^{-1}\int_{0}^{t}\hspace*{-2pt}\int_{t}^{1} L_{n-1}(\gamma)L_{n-1}(\gamma')
h(\gamma)h(\gamma')
\\
&&{}\qquad\hspace*{60pt}\times[F_{n}(\gamma')-F_{n}(\gamma)]\,d\gamma' \,d\gamma
\\
&&\quad= A^{-1} \int_{0}^{t}\hspace*{-2pt}\int_{t}^{1} L_{n-1}(\gamma)L_{n-1}(\gamma')
h(\gamma)h(\gamma')
\\
&&{}\qquad\hspace*{47pt}\times [F_{n}(\gamma')-F_{n}(\gamma)]\,d\gamma'\, d\gamma
\leq0,
\end{eqnarray*}
 since $F_{n}(\gamma)$ is nonincreasing in $\gamma$. Hence,
$\Pi
_{n}(t) \leq\Pi_{n-1}(t)$, which implies that $\Pi_{n}^{-1}(\alpha)
\geq\Pi_{n-1}^{-1}(\alpha)$, that is, $x_{n+1} \geq x_{n}$. Using a
similar argument, one can show that $\Pi_{n}^{-1}(\alpha) \leq\Pi
_{n-1}^{-1}(\alpha)$ if $Y_{n}=1$. This shows that EWOC is
coherent.
\end{pf*}

\section*{Acknowledgments}

We thank the reviewer and executive editor for valuable comments and
suggestions in writing the manuscript.
Mourad Tighiouart is supported in part by NIH/NCI Grants
Nos 1 P01 CA116676, 1 P30 CA138292-01 and 5 P50 CA128613.


\begin{thebibliography}{10}

\bibitem{Adcock1997}
 \textsc{Adcock, C. J.} (1997).
Sample size determination: A review.
\textit{The Statistician} \textbf{46} 261--283.

\bibitem{Babb2004}
\textsc{Babb, J.}  and \textsc{Rogatko, A.} (2004).
\textit{Contemporary Biostatistical Methods in Clinical Trial}.
\textit{Bayesian Methods for Cancer Phase I Clinical Trials} 1--39.
 Dekker, New York.

\bibitem{Babb1998}
\textsc{ Babb, J., Rogatko, A.}  and \textsc{Zacks, S.} (1998).
Cancer Phase I clinical trials: Efficient dose escalation with
overdose control.
\textit{Stat. Med.} \textbf{17} 1103--1120.

\bibitem{Babb2001}
 \textsc{Babb, J. S.} and \textsc{Rogatko, A.} (2001).
Patient specific dosing in a cancer phase I clinical trial.
\textit{Statist. Med.} \textbf{20} 2079--2090.

\bibitem{Braun2003}
 \textsc{Braun, T. M., Levine, J. E.}  and \textsc{Ferrara, J. L. M.} (2003).
Determining a maximum tolerated cumulative dose: Dose reassignment
within the TITE-CRM.
\textit{Controlled Clinical Trials} \textbf{24} 669--681.

\bibitem{Cheng2004}
 \textsc{Cheng, J. D., Babb, J. S., Langer, C., Aamdal,~S.,
Robert, F.}, \textsc{Engelhardt, L. R., Fernberg, O., Schiller, J., Forsberg, G., Alpaugh, R. K., Weiner, L. M.}
and \textsc{Rogatko, A.}  (2004).
Individualized patient dosing in phase I clinical trials: The role of
EWOC in PNU-214936.
\textit{J. Clin. Oncol.} \textbf{22} 602--609.

\bibitem{Cheung2005}
 \textsc{Cheung, Y. K.} (2005).
Coherence principles in dose-finding studies.
\textit{Biometrika} \textbf{92} 863--873.
\MR{2234191}

\bibitem{Cheung2000}
 \textsc{Cheung, Y. K.} and \textsc{Chappell, R.} (2000).
Sequential designs for phase I clinical trials with late-onset
toxicities.
\textit{Biometrics} \textbf{56} 1177--1182.
\MR{1815616}

\bibitem{Chu2009}
 \textsc{Chu, P. L., Lin, Y.} and \textsc{Shih, W. J.} (2009).
Unifying CRM and EWOC designs for phase I cancer clinical trials.
\textit{J. Statist. Plann. Inference} \textbf{139} 1146--1163.
\MR{2479856}


\bibitem{Cox1972}
\textsc{Cox, D. R.} (1972).
Regression models and life-tables.
\textit{J. R. Stat. Soc. Ser. B Stat. Methodol.}
\textbf{34} 187--220.
\MR{0341758}

\bibitem{Dillman1992}
 \textsc{Dillman, R. O.} and \textsc{Koziol, J. A.} (1992).
Phase I cancer trials: Limitations and implications.
\textit{Molecular Biotherapy} \textbf{4} 117--121.

\bibitem{Faries1994}
 \textsc{Faries, D.} (1994).
Practical modifications of the continual reassessment method for
phase I cancer clinical trials.
\textit{J. Biopharm. Statist.} \textbf{4} 147--164.

\bibitem{Garrett-Mayer2006}
 \textsc{Garrett-Mayer, E.} (2006).
The continual reassessment method for dose-finding studies: A
tutorial.
\textit{Clinical Trials} \textbf{3} 57--71.

\bibitem{Gatsonis1992}
 \textsc{Gatsonis, C.} and \textsc{Greenhouse, J. B.} (1992).
Bayesian methods for phase I clinical trials.
\textit{Stat. Med.} \textbf{11} 1377--1389.

\bibitem{Goodman1995}
 \textsc{Goodman, S. N., Zahurak, M. L.} and \textsc{Piantadosi,~S.} (1995).
Some practical improvements in the continual reassessment method for
phase I studies.
\textit{Stat. Med.} \textbf{14} 1149--1161.

\bibitem{Ivanova2006}
 \textsc{Ivanova, A.} (2006).
Escalation, group and A $+$ B designs for dose-finding trials.
\textit{Stat.  Med.} \textbf{25} 3668--3678.
\MR{2252418}

\bibitem{Lin2001}
 \textsc{Lin, Y.} and \textsc{Shih, W. J.} (2001).
Statistical properties of the traditional algorithm-based designs for
phase I cancer clinical trials.
\textit{Biostatistics} \textbf{2} 203--215.

\bibitem{Lunn2000}
 \textsc{Lunn, D. J., Thomas, A., Best, N.} and \textsc{Spiegelhalter, D.} (2000).
WinBUGS---a Bayesian modelling framework: Concepts, structure, and
extensibility.
\textit{Statist. Comput.} \textbf{10} 325--337.

\bibitem{Moller1995}
 \textsc{Moller, S.} (1995).
An extension of the continual reassessment methods using a
preliminary up-and-down design in a dose finding study in cancer
patients, in
order to investigate a greater range of doses.
\textit{Stat. Med.} \textbf{14} 911--922.

\bibitem{NCI2003}
\textsc{NCI} (2003).
Common toxicity criteria for adverse events v3.0 (CTCAE).


\bibitem{OQuigley1990}
 \textsc{O'Quigley, J., Pepe,  M.} and \textsc{Fisher, L.} (1990).
Continual reassessment method: A practical design for phase 1
clinical trials in cancer.
\textit{Biometrics} \textbf{46} 33--48.
\MR{1059105}

\bibitem{OQuigley1999}
 \textsc{O'Quigley, J., Shen, L. Z.} and \textsc{Gamst, A.} (1999).
Two-sample continual reassessment method.
\textit{J. Biopharm. Statist.} \textbf{9} 17--44.

\bibitem{Piantadosi1998}
 \textsc{Piantadosi, S., Fisher, J. D.} and \textsc{Grossman, S.} (1998).
Practical implementation of a modified continual reassessment method
for dose-finding trials.
\textit{Cancer Chemother. Pharmacol.} \textbf{41} 429--436.

\bibitem{Roberts2004}
 \textsc{Roberts, T. G. J., Goulart,  B., Squitieri, L., Stallings, S.~C.,
 Halpern, E. F., Chabner, B. A., Gazelle, G. S., Finkelstein, S.~N.}
and  \textsc{Clark, J. W.} (2004).
Trends in the risks and benefits to patients with cancer
participating in phase 1 clinical trials.
\textit{J.~Amer. Med. Assoc.} \textbf{292} 2130--2140.\

\bibitem{Rogatko2008}
 \textsc{Rogatko, A., Ghosh, P., Vidakovic, B.} and \textsc{Tighiouart, M.} (2008).
Patient-specific dose adjustment in the cancer clinical trial
setting.
\textit{Pharm. Med.} \textbf{22} 345--350.

\bibitem{Rogatko2005}
 \textsc{Rogatko, A., Tighiouart, M.} and \textsc{Xu, Z.} (2005).
EWOC 2.1 application software. Available at
\url{http://sisyphus.emory.edu/software.php}.


\bibitem{Rosenberger2002}
 \textsc{Rosenberger, W. F.} and \textsc{Haines, L. M.} (2002).
Competing designs for phase I clinical trials: A review.
\textit{Stat. Med.} \textbf{21} 2757--2770.

\bibitem{Storer2001}
 \textsc{Storer, B. E.} (2001).
An evaluation of phase I clinical trial designs in the continuous
dose-response setting.
\textit{Stat. Med.} \textbf{20} 2399--2408.

\bibitem{Tighiouart2006a}
 \textsc{Tighiouart, M.} and \textsc{Rogatko, A.} (2006).
\textit{Dose Finding in Drug Development.  Dose Finding in
Oncology--Parametric Methods}  59--72.
Springer, New York.

\bibitem{Tighiouart2006}
 \textsc{Tighiouart, M.} and \textsc{Rogatko, A.} (2006).
\textit{Statistical Methods for Dose-Finding Experiments.  Dose
Escalation with Overdose Control}  173--188.
 Wiley, New York.
\MR{2274231}

\bibitem{Tighiouart2005}
 \textsc{Tighiouart, M., Rogatko, A.} and \textsc{Babb, J. S.} (2005).
Flexible Bayesian methods for cancer phase I clinical trials. Dose
escalation with overdose control.
\textit{Stat. Med.} \textbf{24} 2183--2196.
\MR{2146926}

\bibitem{Tighiouart2007}
 \textsc{Tighiouart, M., Rogatko, A.} and \textsc{Xu, Z.} (2007).
Incorporating patient's characteristics in cancer phase I clinical
trials using escalation with overdose control.
 In \textit{Joint Statistical Meetings}. Salt Lake City.

\bibitem{Ting2006}
 \textsc{Ting, N.} (2006).
\textit{Dose Finding in Drug Development}.
Springer, New York.

\bibitem{Whitehead1997}
 \textsc{Whitehead, J.} (1997).
Bayesian decision procedures with application to dose-finding
studies.
\textit{Int. J. Pharm. Med.} \textbf{11} 201--208.


\bibitem{Xu2007}
 \textsc{Xu, Z., Tighiouart, M.} and \textsc{Rogatko, A.} (2007).
EWOC 2.1: Interactive software for dose escalation in cancer phase I
clinical trials.
\textit{Drug Inform. J.} \textbf{41} 221--228.

\bibitem{Zacks1998}
 \textsc{Zacks, S., Rogatko, A.} and \textsc{Babb, J.} (1998).
Optimal Bayesian-feasibile dose escalation for cancer phase I trials.
\textit{Statist. Probab. Lett.} \textbf{38} 215--220.
\MR{1629891}

\bibitem{Zohar2001}
 \textsc{Zohar, S.} and \textsc{Chevret, S.} (2001).
The continual reassessment method: Comparison of Bayesian stopping
rules for dose-ranging studies.
\textit{Stat. Med.} \textbf{20} 2827--2843.

\bibitem{Zohar2003}
 \textsc{Zohar, S., Latouche, A., Taconner, M.} and \textsc{Chevret, S.} (2003).
Software to compute and conduct sequential Bayesian phase I or II
dose-ranging clinical trials with stopping rules.
\textit{Comput. Methods  Programs  Biomed.} \textbf{72} 117--125.\vspace*{-1pt}


\end{thebibliography}
\end{document}